\documentclass[aps,prc,floatfix,groupedaddress,showpacs,amsfonts,nofootinbib,twocolumn]{revtex4}
\usepackage{bm}
\usepackage{epsfig}
\usepackage{amsmath}
\usepackage{dcolumn}

\begin{document}

\title{Low-energy R-matrix fits for the $^{6}$Li${\bm(d,\alpha)^{4}}$He $S$ factor}

\author{J. Grineviciute}
\affiliation{Dipartimento di Fisica e Astronomia, Universit{\'a} di Catania, I-95125 Catania, Italy}

\author{A. M. Mukhamedzhanov}
\affiliation{Cyclotron Institute, Texas A$\&$M University, College Station, Texas 77843, USA}

\author{L. Lamia}
\affiliation{Dipartimento di Fisica e Astronomia, Universit{\'a} di Catania, I-95125 Catania, Italy}

\author{C. Spitaleri}
\affiliation{Dipartimento di Fisica e Astronomia, Universit{\'a} di Catania, I-95125 Catania, Italy}
\affiliation{Laboratori Nazionali del Sud-INFN, I-95125 Catania, Italy}

\author{M. La Cognata}
\affiliation{Laboratori Nazionali del Sud-INFN, I-95125 Catania, Italy}

\begin{abstract}
\begin{description}
\item[Background:] The information about the $^{6}$Li${(d,\alpha)^{4}}$He reaction rates of astrophysical interest can be obtained by extrapolating direct data to  lower energies, or by indirect methods. The indirect Trojan horse method, as well as various R-matrix and polynomial fits to direct data, estimate the electron screening energies much larger than the adiabatic limit. Calculations that include the subthreshold resonance estimate smaller screening energies.

\item[Purpose:] Obtain the $^{6}$Li${\rm (d,\alpha)^{4}}$He reaction R-matrix parameters and the bare astrophysical $S$ factor for the energies relevant to the stellar plasmas by fitting the R-matrix formulas for the subthreshold resonances to the $S$-factor data above 60 keV.

\item[Methods:] The bare $S$ factor is calculated using the single- and the two-level R-matrix formulas for the closest to the threshold $0^{+}$ and $2^{+}$ subthreshold states at $22.2$, $20.2$ and $20.1$ MeV. The electron screening potential $U{\rm _{e}}$ is then obtained by fitting it as a single parameter to the low-energy data. The calculations are also done by fitting $U{\rm _{e}}$ simultaneously with other parameters.

\item[Results:]  The low-energy $S$ factor is dominated by the $2^{+}$ subthreshold resonance at $22.2$ MeV. The influence of the other two subthreshold states is small. The resultant electron screening is smaller than the adiabatic value. The fits that neglect the electron screening above 60 keV produce a significantly smaller   electron screening potential. The calculations show a large ambiguity associated with a choice of the initial channel radius.

\item[Conclusions:] The R-matrix fits do not show a significantly larger $U_{e}$ than predicted by the atomic physics models. The R-matrix best fit provides $U{\rm _{e}}=149.5$ eV and $S_{b}(0)=21.7$ MeV~b.
\end{description}
\end{abstract}



\maketitle

\section{Introduction}

To obtain the reaction rates relevant for the nuclear astrophysics, experimental data should be extrapolated to the very low-energy region (the Gamow window). The cross section  depends strongly on the energy and therefore is expressed in terms of the astrophysical $S(E)$ factor
\begin{equation}
\sigma\left(E\right) = S\left(E\right)E^{-1} \exp\left(-2 \pi \eta\right).
\label{crSfactor1}
\end{equation}
In the system of units, in which ${\hbar}=c=1$, the Coulomb (Sommerfeld) parameter $\eta=Z_{1}Z_{2}e^{2}\mu/k$. $Z{\rm _{i}}$ is the charge of the nucleus $i$, $k = \sqrt{2\,\mu\,E}$ and $\mu$ are the relative momentum and the reduced mass of the interacting nuclei, $E$ is their relative kinetic energy in the c.m. frame. 

The extrapolation of the cross section down to low energies assumes, that the Coulomb potential of the target nucleus and a projectile results from bare nuclei. In  experimental conditions, however, the Coulomb potential is screened by electrons surrounding the target nucleus, thus reducing the height of the Coulomb barrier and leading to a higher cross section. As the energy approaches zero, the electron screening potential $U_{e}$  enhances the bare nucleus astrophysical factor $ S(E)=S_{b}(E)\exp(\pi\eta U_{e}/E)$. 

The estimated electron screening potential for the $^{6}$Li${\rm (d,\alpha)^{4}}$He reaction in the adiabatic limit is a difference in atomic binding energies between Li and Be$^{+}$, that is 186 eV \cite{ALR57,Sho93}. The Trojan horse (TH) experiment has indirectly measured the bare nucleus astrophysical factor \cite{Spi01}. A comparison of the TH results and a direct measurement led to $U_{e}=340\pm51$ eV \cite{Spi01}. Large $U_{e}$ values were obtained by Engstler et al. \cite{Eng92} using the polynomial fits, as well as by Barker \cite{Ba02} using the R-matrix fits. Barker also noted \cite{Ba02}, that fixing $U_{e}$ at 175 eV results in a reasonable fit with only slightly higher $\chi^{2}$. Ruprecht et al. \cite{Ru04} reported $U_{e}=190\pm50$ eV and concluded, that the lower screening energy is due to the influence of the $2^{+}$ subthreshold resonance \footnote{The subthreshold resonance is defined as a state, which is bound in the entry channel, and is a resonance in the exit channel.}.

The current work presents a new R-matrix analysis for the low-energy $^{6}$Li${\rm (d,\alpha)^{4}}$He reaction, that considers three subthreshold resonances. The largest contribution to the low-energy $S(E)$ factor comes from the resonances and the subthreshold resonances closest to the threshold of the compound nucleus. The three subthreshold resonances closest to the threshold of  $^{8}$Be are the $2^{+};0$ subthreshold state at $-80$ keV, followed by the $0^{+};0$ subthreshold resonance at $-2.08$ MeV and a $2^{+};0$ state at $-2.18$ MeV. The $2^{+};0$ resonance at $2.92$ MeV is not included in the calculation.  The astrophysical $S$ factor is first calculated including only the $2^{+}$ subthreshold resonance at $-80$ keV, and then by adding two more  subthreshold resonances.

\section{Conditions}
\subsection{Experimental data}
The fitting to the direct low-energy experimental data \cite{Eng92,Elw77,Cai85,Ri77,Je62} is done using the nonlinear least-squares procedure. Golovkov et al. \cite{Go81} data are the outliers, and, hence, are not included in the calculation. 

The errors of the experimental data of Engstler et al. \cite{Eng92} do not include the reported uncertainties arising from the number of counts, angular distributions, the effective energy and the target stoichiometry, and, hence, are reduced by $5-10\%$. Including the latter errors would result in a normalized $\chi^{2}_{n} \ll 1$. Here, the normalized $\chi^{2}_{n}$ is defined as $\chi^{2}_{n}=\chi^{2}/\left(N-n_{p}\right)$, where $N$ is the number of experimental points used in the fit calculation, and $n_{p}$ is the number of parameters. 

The error bars of Elwyn et al. experimental data \cite{Elw77} are enlarged to get $\chi^{2}_{n} \approx 1$, as the underestimated errors may lead to a bias in the derived slope. The chosen error bars for Elwyn et al. data \cite{Elw77} are set to the $10 \%$ of the measured $S$-factor value.

\subsection{Parameters}
The R-matrix fits are calculated using the modified R-matrix formulas \cite{Br02}, which use the ``observed'' rather than the formal parameters. The alternate parametrization allows to set the resonances' energies at the experimental values, instead of conventionally \cite{Ba02,LaTh58} choosing the random formal energies in the vicinity of the resonance, and calculating the resulting ``observed'' energies after the R-matrix fit. 

In the three-level fit, the two $\,2^{+}\,$ resonances interfere, and the two-level R-matrix equations are used. The contribution of the $\,0^{+}$ state is  added incoherently. Considering only two channels for each state, the low-energy astrophysical $S(E)$ factor is defined as \cite{LaTh58}
\begin{equation}
S\left(E\right) = \frac{\pi}{2\mu} \exp\left(2 \pi \eta\right) \left( g_{0^{+}}\left|U^{0^{+}}_{cc^{\prime}}\right|^{2}+g_{2^{+}}\left|U^{2^{+}}_{cc^{\prime}}\right|^{2} \right).
\end{equation}
Here, the statistical spin factor $g_{J\pi}$ is
\begin{equation}
g_{J\pi} = \frac{2J+1}{\left(2J_{c}+1\right)\left(2J_{c^{\prime}}+1\right)}.
\end{equation}
The collision matrix $U_{cc^{\prime}}$ in terms of the ``observed'' parameters is identical to the one, that is expressed in terms of the formal parameters \cite{Br02}
\begin{equation}
U_{cc^{\prime}} = \Omega_{c}\Omega_{c^{\prime}}\left( \delta_{cc^{\prime}} +2i \left(P_{c^{\prime}}P_{c}\right)^{1/2}\gamma^{T}_{c^{\prime}} A \gamma_{c} \right),
\end{equation}
and the $A^{-1}$ matrix in terms of the ``observed'' parameters is defined as
\begin{eqnarray}
\left(A^{-1}\right)_{ij}= \left(E_{i}-E\right)\delta_{ij}-\sum_{c} \gamma_{ic} \gamma_{jc} \left(S_{c}+iP_{c}\right) \nonumber\\
+\sum_{c} \left\{
\begin{array}{ll}
	\gamma^{2}_{ic} S_{ic} & i=j \\
	\gamma_{ic}\gamma_{jc} \frac{S_{ic}\left(E-E_{j}\right)-S_{jc}\left(E-E_{i}\right)}{E_{i}-E_{j}} & i\neq j \; .
\end{array}
\right.
\end{eqnarray}
Here, $S_{ic}$ is the shift function $S_{c}$, evaluated at energy $E_{i}$, $P_{c}$ is the penetribility, and $\Omega_{c}=\left(I_{c}/O_{c}\right)^{1/2}$.

The R-matrix formula of Lane and Thomas \cite{LaTh58} requires an inclusion of all partial waves. For the initial $2^{+}$ state one would have $\,(l = 0,\, s = 2)$, $\,(l = 2,\, s = 0)$, $\,(l = 2,\, s = 1)$, $\,(l = 2,\, s = 2)\,$ partial waves and $\,(l = 0,\, s = 0)$, $\,(l = 2,\, s = 2)\,$ partial waves for the initial $\,0^{+}\,$ state.  The exit channel for the $\,2^{+}\,$ state is $\,(l = 2, s = 0)\,$, and, for the $\,0^{+}\,$ state, it is $\,(l = 0, s = 0)$. The asymptotic normalization coefficients (ANC's) for the bound states, as well as the reduced width amplitudes $\gamma_{\alpha}$'s for the final channels are unknown and, therefore, are treated as parameters in the R-matrix calculation. Assuming that the $\,s\,$-wave approximation for the incoming deuteron is reasonable at very low deuteron energies, the number of free parameters reduces significantly.

\subsection{Constraints}

Unconstrained fits to direct data do not allow reliable determination of all parameters; therefore present calculations use constrained fits, which are subject to assumptions of the physical parameters. First, the R-matrix fits depend on the restrictions placed on the experimentally unknown $\alpha + \alpha$ channels' partial widths. 

The subthreshold resonances under consideration are broad. The experimental resonance widths  of  $\,{}^{8}{\rm Be}$ are 0.8 MeV and 0.88 MeV for the $\,2^{+}\,$ states at 22.2 MeV and 20.1 MeV, respectively, and 0.72 MeV for the $\,0^{+}\,$ state at 20.2 MeV \cite{Ti04}. The total width for each level is a sum of all partial widths. The threshold of the $\,{}^{6}$Li${(d,\alpha)^{4}}$He reaction entrance channel corresponds to a high excitation energy in a compound nucleus, hence, many reaction channels are open. Considering only the $\,{}^{6}$Li${\rm (d,\alpha)^{4}}$He reaction, the major contribution to the total width at the subthreshold energy comes from the $\,\alpha +\alpha\,$ channel due to the large Q value of the reaction $($Q$ = 22.37$ MeV$)$ and varies slowly with energy. The relative widths  $\,\frac{\Gamma\left(\alpha\right)}{\Gamma\left(p\right)}\,$ of the $\,2^{+}\,$ subthreshold resonance at $20.1$ MeV have been determined experimentally to be $\,4.5 \pm 0.6\,$ \cite{Pu92}. $\,\Gamma_{\alpha}/\Gamma < 0.5$ has been reported \cite{Be67} for the $0^{+}$ state. 

Page \cite{Pa05} performed a many-level multichannel simultanous R-matrix fit to known $\alpha + \alpha$ elastic scattering data, as well as $(\rm d,\alpha)$ and other channels' data. However, a single-level fit to data, as well as the fit, that includes all three aforementioned subthreshold resonances, fails for the bare $^{6}$Li${\rm (d,\alpha)^{4}}$He  astrophysical $S$ factor using the suggested formal $\alpha + \alpha$ channels partial widths $\,\Gamma_{\alpha}=0.11\,$ MeV for the $\,2^{+}\,$ subthreshold state at 22.2 MeV (experimental total width 0.8 MeV), $\,\Gamma_{\alpha}=0.55\,$ MeV for the $\,0^{+}\,$ subthreshold state  at 20.2 MeV (experimental total width 0.72 MeV), and $\,\Gamma_{\alpha}=0.17\,$ MeV for the $\,2^{+}\,$ subthreshold state at 20.1 MeV (experimental total width 0.88 MeV) in Ref. \cite{Pa05}. The fit also fails, if one includes an arbitrary background.

Reference \cite{Mu10} notes that because the resonances are broad, the resonance contribution cannot be separated from the background contribution  and therefore the elastic scattering data do not provide accurate resonance parameters. Also, the best fit of Ref. \cite{Pa05} includes an additional unknown level at 580 keV (excitation energy 22.78 MeV) with the formal $\,\Gamma_{\alpha}=0.04\,$ MeV and $\,\Gamma_{\rm d}=0.23\,$ MeV for the $s$-wave deuteron.

The relative width  $\,\Gamma\left(\alpha\right) / \Gamma\left(p\right)\,$ for the  $2^{+}$ subthreshold resonance at $20.1$ MeV, found in \cite{Pa05}, is significantly smaller than that given in literature \cite{Pu92} and obtained from the $p +^{7}$Li$\rightarrow\alpha + \alpha$ and $n +^{7}$Be$\rightarrow\alpha + \alpha$ reactions data. Also, the $\,\Gamma_{\alpha}$ of the $2^{+}$ subthreshold resonance at $22.2$ MeV obtained by Ref. \cite{Pa05} is much smaller than that obtained by other R-matrix or polynomial fits to data \cite{Ba02,Ru04,Eng92}. Hence, the present study considers only those constrains, that are imposed by the experimental measurements.

For the bound states the ANC's are related to the reduced width amplitudes \cite{muk99}
\begin{equation}
\gamma^{2}_{c}=\frac{1}{2\mu} \frac{W^{2}_{-\eta_{\kappa_{c}},l+1/2} \left(2\kappa_{c}r_{0}\right)}{r_{0}} \left|C\right|^{2}\; ,
\end{equation}
where $W_{-\eta_{\kappa_{c}},l+1/2} (2\kappa_{c}r_{0})$ is a Whittaker function. Without the experimentally imposed constrains, it is not possible to say which range of the ANC's is more appropriate. Therefore the ANC's are treated as free parameters.

\subsection{Channel radii}

The channel radii associated with the range of the nuclear force are calculated by a conventional formula of Lane and Thomas \cite{LaTh58} $r=r_{0}\left(A^{1/3}_{1}+A^{1/3}_{2}\right)$, where $A_{1}$ and $A_{2}$ are the mass numbers and $r_{0}$ is a numerical value between $1.4$ and $1.5$ fm. 
In principle, the collision matrix is independent of the choice of the channel radii, provided a large enough number of levels is included into the analysis, and, consequentially, is the astrophysical $S$ factor. Partial widths and energies of the resonances resulting from an R-matrix fit calculation should also be channel radii independent \cite{De10}. The sensitivity of the resonance parameters to the adopted channel radii in the initial and final channels is illustrated in Table~\ref{Tab1} for a single-level calculation that uses the $\,s\,$-wave approximation for the deuteron. The parameters depend strongly on the $^{6}$Li+d channel radius, which may indicate a need for the inclusion of the additional initial channel partial waves in the R-matrix fit. Also, as the analysis deals with broad resonances, the strong dependance on channel radii may support a need for an inclusion of additional levels into the R-matrix fit.

\begin{table*}[ht]
\caption{\label{Tab1} The $2^{+}$ subthreshold resonance energy, the partial width of the $\,\alpha +\alpha\,$ channel and the bare $S$ factor resulting from a single-level R-matrix best fit to the low-energy Engstler data \cite{Eng92} above 60 keV (32 data points) for various channel radii. The reduced width amplitudes for both channels and the energy are treated as free parameters. The electron screening $U_{e}$ is then fit as a single parameter to the Engstler data (64 data points).}
\begin{ruledtabular}
\begin{tabular}{ccccccc}
$R_{i}\: {\rm (fm)}$ & $R_{f}\: {\rm (fm)}$ & $E_{2^{+}}{\rm (MeV)}$ & $\Gamma_{\alpha}$\:(MeV) & $S_{b}(0)\: {\rm (MeV \: b)}$ &  $\chi^{2}_{n}$ & $U_{e}$\:(eV) \\ 
\hline\\
4.5 &  4.0 & 22.2610 &  0.6952 & 22.5222 & 0.7678 & 76.9987\\ 
4.5 &  4.5 & 22.2608 &  0.6947 & 22.5279 & 0.7681 & 76.5234\\ 
4.5 &  5.0 & 22.2606 &  0.6943 & 22.5278 & 0.7682 & 76.6777\\ 
4.5 &  5.5 & 22.2605 &  0.6940 & 22.5309 & 0.7684 & 76.4070\\ 
5.0 &  4.0 & 22.1911 &  0.7928 & 22.4021 & 0.7494 & 83.2557\\ 
5.0 &  4.5 & 22.1907 &  0.7923 & 22.4066 & 0.7496 & 82.8810\\ 
5.0 &  5.0 & 22.1904 &  0.7919 & 22.4087 & 0.7497 & 82.7473\\ 
5.0 &  5.5 & 22.1902 &  0.7917 & 22.4115 & 0.7498 & 82.4966\\ 
5.5 &  4.0 & 22.1368 &  0.9330 & 22.2113 & 0.7303 & 92.8859\\ 
5.5 &  4.5 & 22.1362 &  0.9327 & 22.2158 & 0.7305 & 92.4693\\ 
5.5 &  5.0 & 22.1357 &  0.9325 & 22.2155 & 0.7306 & 92.6005\\ 
5.5 &  5.5 & 22.1354 &  0.9323 & 22.2184 & 0.7307 & 92.3145\\ 
\end{tabular}
\end{ruledtabular}
\end{table*}

When allowing the channel radius to vary as one of the parameters, the single-level best fit sets the initial channel radius $R_{i}=6.6$ fm. The best fit places the subthreshold resonance at 22.252 MeV with $\Gamma_{\alpha}=0.2351$ MeV. The corresponding $U_{e}=95.097$ eV. There is no obvious reason, however, to set the initial channel radius to this value, as one looks for a range of radii in which the conclusions of the calculation are reasonably stable, rather than for a single value which produces the lowest $\chi^{2}$. In this analysis, the chosen values for the initial and final channel radii are 5.0 and 4.5 fm, respectively. The sensitivity to the initial channel radius is evaluated in the error bars.

As deviations of the conclusions could be attributed to the effects of other levels, the single-level $R$-matrix fit for the $2^{+}$ subthreshold resonance and the $R$-matrix fit, that includes the three aforementioned subthreshold states, are considered. The calculations use the $\,s\,$-wave approximation for the deuteron, as with the existing data including more partial waves  would only introduce more unknown fitting parameters. 

\section{Results}
\subsection{Single-level fit}

\begin{figure}[ht]
\epsfig{file=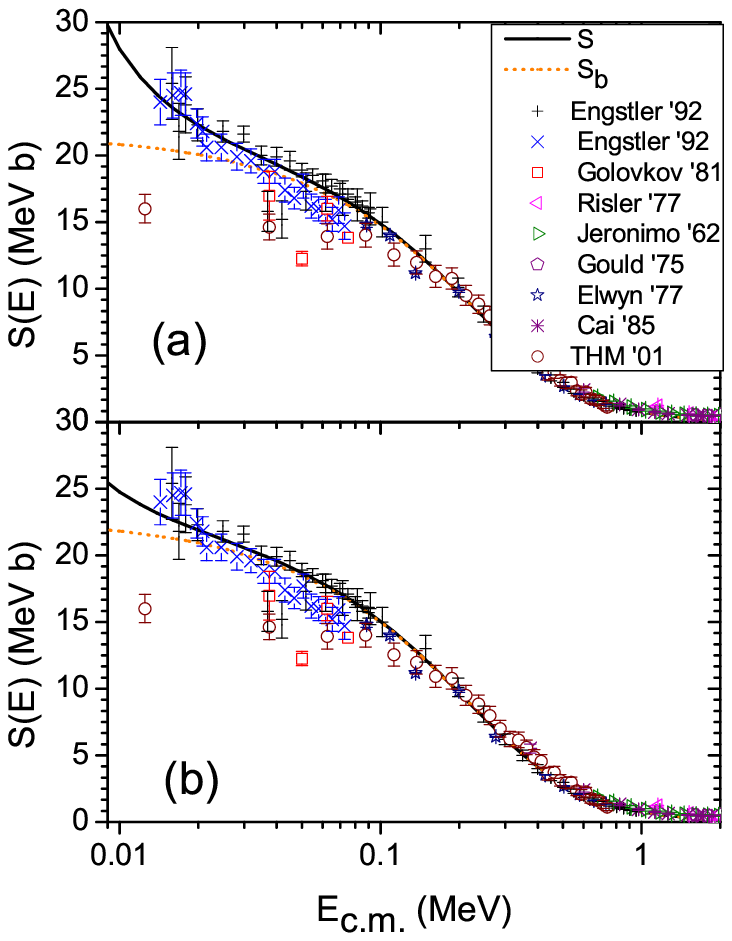,width=8cm}
\caption{\label{FigS1} (Color online) The astrophysical factor $S(E)$ for the $^{6}$Li${(d,\alpha)^{4}}$He reaction, resulting from a single-level $R$-matrix best fit to the experimental data \cite{Eng92,Elw77,Cai85,Ri77,Je62}, as a function of the relative kinetic energy $E$ in the entrance channel. The solid line and the dotted line correspond to the observed and the bare  astrophysical $S$ factor, respectively. The top panel shows the best fit to data below $1$ MeV, that uses four free parameters:  the reduced width amplitudes $\gamma_{\alpha}$ and $\gamma_{d}$, the $2^{+}$ subthreshold resonance energy $E{\rm _{R}}$ near the 22.2 MeV excitation energy  and the electron screening potential $U_{e}$. The bottom panel shows the best fit to data in a 60 keV -- 1 MeV range, that uses $\gamma_{\alpha}$ and $\gamma_{d}$ as the only parameters. The electron screening potential $U_{e}$ is then fit to data \cite{Eng92,Elw77,Cai85,Ri77,Je62} below 1 MeV as a single parameter.  Engstler et al. data \cite{Eng92} are shown as pluses for atomic target and crosses for molecular target. The circles are the bare $S_{b}(E)$ factor from Trojan horse experiment \cite{Spi01}. Other data \cite{Elw77,Je62,Ri77,Cai85,Go81,GJB75} are taken from \cite{exfor}. The errors of Engstler et al. data \cite{Eng92} do not include reported uncertainties arising from number of counts, angular distributions, effective energy and target stoichiometry and hence are reduced by $5-10\%$.  The error bars of experimental data of Elwyn et al. \cite{Elw77} are shown as in \cite{exfor}, but are enlarged in the calculation.}
\end{figure}

An unrestricted single-level $R$-matrix best fit to the low-energy experimental data \cite{Eng92,Elw77,Cai85,Ri77,Je62} that uses the energy $E_{R}$ of a $2^{+}$ subthreshold resonance near the 22.2 MeV excitation energy, the electron screening potential $U_{e}$, as well as the reduced width amplitudes $\gamma_{\alpha}$ and $\gamma_{d}$ as free parameters, is shown in the top panel of Fig.~\ref{FigS1}. The best fit places the subthreshold resonance at $22.1692^{+0.0775}_{-0.0625}$ MeV with $\,\Gamma_{\alpha}=0.8378^{+0.1564}_{-0.1044}\,$ MeV.  The resulting bare astrophysical $S$ factor at zero energy $S_{b}(0) = 21.6144^{+0.0725}_{-0.0573}$ MeV~b and the electron screening $U_{e} = 161.953^{+7.648}_{-4.557}$ eV. The $\chi^{2}$ of the fit, that uses four parameters and 82 data points, is $118.61$. The normalized $\chi^{2}_{n}$ is then $1.5207$.

The subthreshold energy of the $2^{+}$ state is then fixed at the experimental value of $-80$ keV (excitation energy 22.2 MeV) and the partial width of the $\,\alpha +\alpha\,$ channel is restricted not to exceed the total width of the state. The best fit that treats $U_{e}$, $\gamma_{\alpha}$ and $\gamma_{d}$ as parameters, provides $S_{b}(0)=22.1571^{+0.0322}_{-0.5271}$ MeV~b, $\,\Gamma_{\alpha}=0.5132^{+0.1952}_{-0.1141}\,$ MeV and the electron screening $U_{e}=126.665^{+43.231}_{-2.807}$ eV. The $\chi^{2}=119.69$ is only slightly higher than that resulting from the former fit, in which the subthreshold resonance energy is allowed to vary, resulting in a slightly smaller $\chi^{2}_{n}=1.5147$.

The ambiguity of the fit due to the sensitivity to the choice of the channel radii is larger for the restricted fit. The larger errors correspond to the fits for which the initial channel radius was reduced, resulting in larger values of the electron screening, the larger $\alpha $ partial width and the smaller $S_{b}(0)$.

To neglect the electron screening, the single-level R matrix is fit to data above $60$ keV \cite{Ru04}. The calculation for the bare $S$ factor is done treating $\gamma_{\alpha}$ and $\gamma_{d}$ as the only parameters. The observed $S$ factor is then obtained by varying $U_{e}$ as a single parameter to fit data below 1 MeV \cite{Eng92,Elw77,Cai85,Ri77,Je62}. The best fit is shown in the bottom panel of Fig.~\ref{FigS1}. Resulting $S_{b}(0)=22.7446^{+0.0104}_{-1.0660}$ MeV~b and $\,\Gamma_{\alpha}=0.5244^{+0.1864}_{-0.1212}\,$ MeV. $\chi^{2}$ of the best fit to 50 data points is 66.6727, $\chi^{2}_{n}=1.3890$. The reduced width amplitudes are $\gamma_{\alpha}=-0.2105$ MeV$^{-1/2}$ and $\gamma_{d}=0.4341$ MeV$^{-1/2}$. The corresponding ANC is $C=3.1449$ fm$^{-1/2}$.  Resultant $U_{e}=69.288^{+98.537}_{-0.678}$ eV. 

The sensitivity for the latter fit tracks the same way as the previous fit: Reducing the channel radius strongly increases the electron screening. Also, the single-level $R$-matrix best fit at a fixed energy of the $2^{+}$ subthreshold resonance at -80 keV, results in different sets of parameters, when fitting the electron screening simultaneously with other parameters, and, when the electron screening is neglected below 60 keV. Fitting all data below 1 MeV with the screening potential $U_{e}$ included in the set of the fitting parameters, results in $U_{e}= 126.7$ eV, while fitting data at energies above 60 keV, where supposedly the electron screening can be neglected, and then using the $U_{e}$ as a single fitting parameter to fit all data below 1 MeV, results in the nearly twice as low screening potential $U_{e}= 69.3$ eV.

\subsection{Two-level fit}
The second closest subthreshold resonance to the threshold of $^{8}$Be (excitation energy 22.28 MeV) is the $0^{+};0$ state at -2.08 MeV (excitation energy 20.2 MeV). The $S$ factors for the two states are added incoherently. 

The $R$-matrix best fit, that treats the $\gamma_{\alpha}$'s and the ANC's as free parameters, results in a much too large width of the $0^{+};0$ state. Restricting the partial widths of the $\alpha + \alpha$ channels not to exceed the total widths of each state leads to $S_{b}(0)=22.9535^{+0.0054}_{-1.1700}$ MeV~b,  $\,\Gamma_{2^{+}}=0.5070^{+0.1883}_{-0.1186}\,$ MeV, $\,\Gamma_{0^{+}}=0.7115^{+0.0816}_{-0.5849}\,$ MeV.  The best fit parameters are ANC $=3.0533$ fm$^{-1/2}$ and $\gamma_{\alpha}=-0.2069$ MeV$^{-1/2}$ for the $2^{+}$, and ANC $=12.6073$ fm$^{-1/2}$ and $\gamma_{\alpha}=0.242$ MeV$^{-1/2}$ for the $0^{+}$ state. $\chi^{2}$ of the fit, that uses four parameters and 50 data points, is 65.8849, $\chi^{2}_{n}=1.4323$. The electron screening is then fit to data below 1 MeV as a single parameter, resulting in $U_{e}=56.477^{+103.391}_{-0.200}$ eV. 

By restricting the $\Gamma_{\alpha}$ of the $0^{+};0$ subthreshold resonance not to exceed the half of the total width, as reported in \cite{Be67}, the best fit parameters become ANC $=3.1437$ fm$^{-1/2}$ and $\gamma_{\alpha}=-0.2104$ MeV$^{-1/2}$ for the $2^{+}$, and ANC $=1.1062$ fm$^{-1/2}$ and $\gamma_{\alpha}=0.0823$ MeV$^{-1/2}$ for the $0^{+}$ state. $\chi^{2}$ of the best fit is 66.6689 and $\chi^{2}_{n}= 1.4493$. Then $S_{b}(0)=22.7512^{+0.1075}_{-1.0184}$ MeV~b, $\,\Gamma_{2^{+}}=0.5243^{+0.1786}_{-0.1213}\,$ MeV, $\,\Gamma_{0^{+}}=0.0823^{+0.3084}_{-0.0325}\,$ MeV, and the electron screening  $U_{e}$ is $68.670^{+95.031}_{-1.134}$ eV. 

Thus, adding the second subthreshold resonance hardly affects the extracted screening potential from that obtained for a single subthreshold state. It also does not reduce the sensitivity to the channel radii. One notes that whether a single  $2^{+}$ subthreshold state is used in the $R$-matrix fit, or the two subthreshold states $2^{+}$ and $0^{+}$ are considered, the determined alpha partial width of the $2^{+}$ state is significantly larger than the one obtained in Ref. \cite{Pa05}.

\subsection{Three-level fit} 

In this section, the $R$-matrix fit considers three subthreshold resonances: two $\,2^{+}\,$ and one $\,0^{+}$. The $\,2^{+}\,$ states are interfering and the $\,0^{+}$ state is added incoherently.

The partial widths of the $\alpha + \alpha$ channels for each state are constrained not to exceed the total experimental widths. In the absence of further constrains, the $R$-matrix best fit to the low-energy experimental data \cite{Eng92,Elw77,Cai85,Ri77,Je62} for the bare astrophysical factor between 60 keV and $1$ MeV results in  $S_{b}(0) = 22.6507^{+0.3972}_{-0.8690}$ MeV~b. The best fit parameters are shown in Table \ref{TabPar} as a set 1. The partial widths of the $\alpha + \alpha$ channels are $\,\Gamma_{2^{+}}=0.6932^{+0.0020}_{-0.1867}\,$ MeV, $\,\Gamma_{0^{+}}=0.4892^{+0.2204}_{-0.4790}\,$ MeV and $\,\Gamma_{2^{+}}=0.1462^{+0.0128}_{-0.1451}\,$ MeV for the subthreshold levels at 22.2 MeV, 20.2 MeV and 20.1 MeV respectively. $\chi^{2}$ of the fit, that uses six parameters and 50 data points, is 65.5116, $\chi^{2}_{n}=1.4889$.   The electron screening is then fit to all data below 1 MeV as a single parameter, resulting in $U_{e}=71.393^{+88.7313}_{-19.603}$ eV.  The best fit looks identical to the panel (b) of Fig.~\ref{FigS1}. 

\begin{table*}[ht]
\caption{\label{TabPar}  The $R$-matrix best fit observed parameters for three subthreshold resonances to experimental data \cite{Eng92,Elw77,Cai85,Ri77,Je62}. The first parameter set results from the best fit to data between 60 keV and 1 MeV (50 data points) treating the reduced width amplitudes $\gamma_{\alpha}$'s and the asymptotic normalization coefficients ANC's as parameters. $U_{e}$ is then fit as a single parameter to all data. The second set results from the fit to data below 1 MeV (82 data points) simultaneously fitting the reduced width amplitudes $\gamma_{\alpha}$'s, ANC's and the electron screening potential $U_{e}$.}
\begin{ruledtabular}
\begin{tabular}{cccccccccc}
 & & \multicolumn{2}{c}{Parameter set 1}& &  & \multicolumn{3}{c}{Parameter set 2}& \\
\cline{3-4} \cline{7-9} 
$J^{\pi}T$  & $E_x {\rm (MeV)}$ & & & & & & & & \\
 & & C (fm$^{-1/2}$) &  $\gamma_{\alpha} ({\rm MeV}^{-1/2})$&  $U_{e}$ (eV)&$\Gamma_{\alpha} \rm{(MeV)}$  &  C (fm$^{-1/2}$) &  $\gamma_{\alpha} ({\rm MeV}^{-1/2})$&  $U_{e}$ (eV)&$\Gamma_{\alpha} \rm{(MeV)}$  \\ 
\hline\\
$2^{+}0$ & 22.2 & 2.2607 &0.2420& & 0.6932&  2.2378 &0.2558& &0.7746   \\ 
$0^{+}0$ & 20.2 & 1.7430 &0.2009& & 0.4892&  1.2790 &0.0366& & 0.0163\\ 
$2^{+}0$ & 20.1 & 2.4390 &0.1146&71.393& 0.1462 &  2.1159 &0.0925& 149.521& 0.0952 \\ 
 & & \multicolumn{2}{c}{ $\chi^{2}_{n}=1.4889$} & & & \multicolumn{3}{c}{ $\chi^{2}_{n}=1.5671$}&\\
\end{tabular}
\end{ruledtabular}
\end{table*}

The sensitivity of the fit to the choice of channel radii for the three level fit tracks the same way as the sensitivity for a single-level fit, possibly indicating that this problem is not due to the lack of the background levels. $\chi^{2}=65.5116$ for the three-level fit is only slightly smaller than the $\chi^{2}$ of a single level fit, producing a slightly higher $\chi^{2}_{n}$. Due to a strong sensitivity to the channel radii, it is difficult to precisely determine the partial widths of the subthreshold resonances; however, it is clear that the fit prefers a partial width of the $2^{+}$ state at 20.1 MeV significantly lower than the experimental total width 0.88 MeV \cite{exfor} and its effect on the $S$ factor is small. 

The $R$-matrix best fit to the low-energy experimental data \cite{Eng92,Elw77,Cai85,Ri77,Je62} that fits the ANC's, the reduced width amplitudes $\gamma_{\alpha}$'s and the electron screening simultaneously, looks very similar to the panel (a) of Fig.~\ref{FigS1}.  S$_{b}(0)$ of the best fit is $21.7186^{+0.5402}_{-0.0580}$ MeV~b and $U_{e}=149.521^{+19.307}_{-30.461}$ eV. The best fit parameters are shown in Table \ref{TabPar} as a set 2. Resulting partial widths of $\alpha + \alpha$ channels are $\,\Gamma_{2^{+}}=0.7746^{+0.0086}_{-0.3690}\,$ MeV, $\,\Gamma_{0^{+}}=0.0163^{+0.4928}_{-0.0}\,$ MeV and $\,\Gamma_{2^{+}}=0.0952^{+0.0013}_{-0.0941}\,$ MeV for the subthreshold levels at 22.2 MeV, 20.2 MeV and 20.1 MeV respectively.   $\chi^{2}$ of the fit, that uses 7 parameters and 82 data points is 117.5280, $\chi^{2}_{n}=1.5671$.   

The electron screening potential $U_{e}=149.521$ eV, resulting from the best fit that varies all parameters simultaneously is halved, when the R matrix is fit to the experimental data above 60 keV to neglect the electron screening. A comparison of the best fit parameters, when $U_{e}$ is fit simultaneously and separately, listed in Table \ref{TabPar}, possibly suggests, that the electron screening is not negligible above 60 keV, contrary to Ref. \cite{Ru04} stating that for the deuteron energies larger than 50 keV the enhancement of the cross section due to the screening effect can be neglected. Therefore, the parameters we recommend are those resulting from a fit that varies $U_{e}$ simultaneously with other parameters. 

The bare $S$ factor resulting from different $R$-matrix fits is larger than many previously reported calculations \cite{Ba02} and agrees with S$_{b}(0)=23 \pm 2.5$ MeV~b obtained by Ruprecht et al. \cite{Ru04}. While the value of the electron screening potential $U_{e}$ is sensitive to the choice of channel radii, it is smaller than the adiabatic limit. We did not observe a significantly larger $U_{e}$, as it was reported in Ref. \cite{Spi01,Eng92,Ba02}.

\section{Conclusion}

The astrophysical $S$ factor for the low-energy  $^{6}$Li${(d,\alpha)^{4}}$He reaction dominated by  broad subthreshold resonances has been analyzed using the single-, two- and three level $R$-matrix fits. For the low-energy R matrix we use the s-wave approximation for the deuteron. The resulting ambiguity due to the choice of channel radii is large in a single-level fit, Table \ref{Tab1}, as well as in the fit that considers three subthreshold resonances. 

Our goal is to check possibility of determination of the electron screening potential from the low-energy astrophysical $S$ factors and to determine the ANC's of the subthreshold states and $\alpha$ partial widths.  We find that parameters depend on the number of the subthreshold states involved in the fitting. We consider the fit with three subthreshold states as the most reliable.  We find that the extracted screening potential depends on the used procedure. If we first fit the $S$ factor varying all the parameters at energies above 60 keV, at which  according to Ref. \cite{Ru04}
the electron screening potential can be neglected, and then fixing all the parameters and varying only $U_{e}$ to fit the astrophysical factor at energies below 1 MeV, we obtain $U_{e} = 71.4$ eV.  However, if we fit the $S$ factor at energies below 1 MeV varying all the parameters simultaneously including $U_{e}$, we get $U_{e}= 149.5$ eV.  Thus, the result strongly depends on the fitting procedure.  We may conclude that the assumption of Ref. \cite{Ru04} that electron screening effects are negligible at energies above 50 keV is not valid and our recommended value is $U_{e}= 149.5$ eV. Note, that other obtained fitting parameters are also sensitive to the fitting procedure. Our recommended values are the parameters from the set 2 in Table \ref{TabPar}.

Our obtained $\alpha$ width for the $\,2^{+}\,$ subthreshold resonance at  22.2 MeV  $\Gamma_{\alpha} = 0.77$ MeV is higher than the 0.56 MeV value obtained by Ref. \cite{Ru04} and is closer to the 0.76 MeV value obtained by \cite{Ba02}. The partial width is lower than the total experimental value of 0.8 MeV \cite{exfor}, which was confirmed by corresponding theoretical calculations. The $\alpha$ partial width for the $\,0^{+}\,$  subthreshold state at 20.2 MeV is small and agrees with  $\,\Gamma_{\alpha}/\Gamma < 0.5$ reported in Ref. \cite{Be67}. The partial width of $\,2^{+}\,$  state at 20.1 MeV is much lower than the experimental value of the total width \cite{exfor}.  Our attempt to fit the $S$ factor for the $^{6}$Li${(d,\alpha)^{4}}$He reaction using $\,\alpha$ widths from Ref. \cite{Pa05} failed and we believe, that the main reason for that is that our $\alpha$ partial width for the dominant $\,2^{+}\,$ state at 22.2 MeV is significantly higher than 0.11 MeV obtained in \cite{Pa05}.  We conclude that while it is difficult to pinpoint accurately the electron screening potential by fitting existing direct measurements, the obtained electron screening potential is definitely smaller than the adiabatic limit  $U_{e}= 186$ eV \cite{ALR57,Sho93}.

\section*{Acknowledgments}
This work has been partially supported by the Italian Ministry of the University under Grant RBFR082838 (FIRB2008). A.~M.~M. acknowledges the support by the US Department of Energy under Grants No. DE-FG02-93ER40773, No. DE-FG52-09NA29467 and NSF under Grant No. PHY-1415656.

\end{document}